\documentclass[aps,prb,twocolumn,superscriptaddress,showpacs,showkeys,floatfix]{revtex4}

\usepackage{graphicx,color}
\usepackage{multirow,slashbox}

\graphicspath{{figs/}}
\begin{document}
\title{Polar Surface Effects on the Thermal Conductivity in ZnO Nanowires: a Shell-Like Surface Reconstruction-Induced Preserving Mechanism}
\author{Jin-Wu Jiang}
    \altaffiliation{Corresponding author: jwjiang5918@hotmail.com}
    \affiliation{Shanghai Institute of Applied Mathematics and Mechanics, Shanghai Key Laboratory of Mechanics in Energy Engineering, Shanghai University, Shanghai 200072, People's Republic of China}
    \affiliation{Institute of Structural Mechanics, Bauhaus-University Weimar, Marienstr. 15, D-99423 Weimar, Germany}
\author{Harold S. Park}
    \altaffiliation{Corresponding author: parkhs@bu.edu}
    \affiliation{Department of Mechanical Engineering, Boston University, Boston, Massachusetts 02215, USA}
\author{Timon Rabczuk}
    \altaffiliation{Corresponding author: timon.rabczuk@uni-weimar.de}
    \affiliation{Institute of Structural Mechanics, Bauhaus-University Weimar, Marienstr. 15, D-99423 Weimar, Germany}
    \affiliation{School of Civil, Environmental and Architectural Engineering, Korea University, Seoul, South Korea }

%\date{22 December 2009}
\date{\today}
\begin{abstract}

We perform molecular dynamics (MD) simulations to investigate the effect of polar surfaces on the thermal transport in zinc oxide (ZnO) nanowires. We find that the thermal conductivity in nanowires with free polar (0001) surfaces is much higher than in nanowires that have been stabilized with reduced charges on the polar (0001) surfaces, and also hexagonal nanowires without any transverse polar surfaces, where the reduced charge model has been proposed as a promising stabilization mechanism for the (0001) polar surfaces for ZnO nanowires.  From normal mode analysis, we show that the higher thermal conductivity is due to a shell-like reconstruction that occurs for the free polar surfaces.  This shell-like reconstruction suppresses twisting motion in the nanowires such that the bending phonon modes are not scattered by the other phonon modes, and leads to substantially higher thermal conductivity in the ZnO nanowire with free polar surfaces. Furthermore, the auto-correlation function of the normal mode coordinate is utilized to extract the phonon lifetime, which leads to a concise explanation for the higher thermal conductivity in ZnO nanowires with free polar surfaces. Our work demonstrates that ZnO nanowires without polar surfaces, which exhibit low thermal conductivity, are more promising candidates for thermoelectric applications than nanowires with polar surfaces (and also high thermal conductivity).
\end{abstract}

\maketitle
\pagebreak

\section{introduction}
Zinc oxide (ZnO) nanowires (NWs) are wide bandgap semiconductors with unique piezoelectric and optical properties.\cite{YangP2002,PeartonSJ,WangZL,OzgurU,WollC,WangZL2012}.  ZnO NWs have been found, primarily due to surface effects resulting from their large surface to volume ratio, to have substantially different properties from their bulk counterparts.  These surface effects become even more important in ZnO NWs if the surfaces are polar (0001) surfaces.  Hence, much research has focused on investigating surface effects on various physical and chemical properties of the ZnO NWs, including the mechanical properties,\cite{KucheyevSO,ChenCQ,ZhangL2006apl,KulkarniAJ,AgrawalR} the electronic or optical band structure,\cite{LinKF,LinKF2006apl,LanyS} the piezoelectric properties,\cite{MitrushchenkovA,DaiS2013jmps} and others.

The (0001) surfaces are particularly important for ZnO NWs because they are polar.  If ZnO is truncated in the (0001) and (000$\bar{1}$) directions, a Zn ion layer will be exposed at one end while an O ion layer will be exposed at the opposite surface.\cite{WanderA} The two exposed surfaces are strongly polarized, and thus are unstable unless stabilized through one of several possible mechanisms.  In clean environments, the polar surfaces prefer a shell-like reconstruction by outward relaxation of the outermost ion layers.\cite{JedrecyN,KulkarniAJ2005} In another stabilization mechanism, the polar (0001) surfaces can become stable with the saturation of the surface bonds.\cite{WanderA,CarlssonJM,LauritsenJV} A third stabilization mechanism is to transfer some negative charge from the (000$\bar{1}$)-O polar surface to the (0001)-Zn polar surface, which results in the reduction of the charges on the polar surfaces by a factor of 0.75.\cite{NogueraC,DaiS2011jap,DaiS2013jmps}

\begin{figure*}[htpb]
  \begin{center}
    \scalebox{1.0}[1.0]{\includegraphics[width=\textwidth]{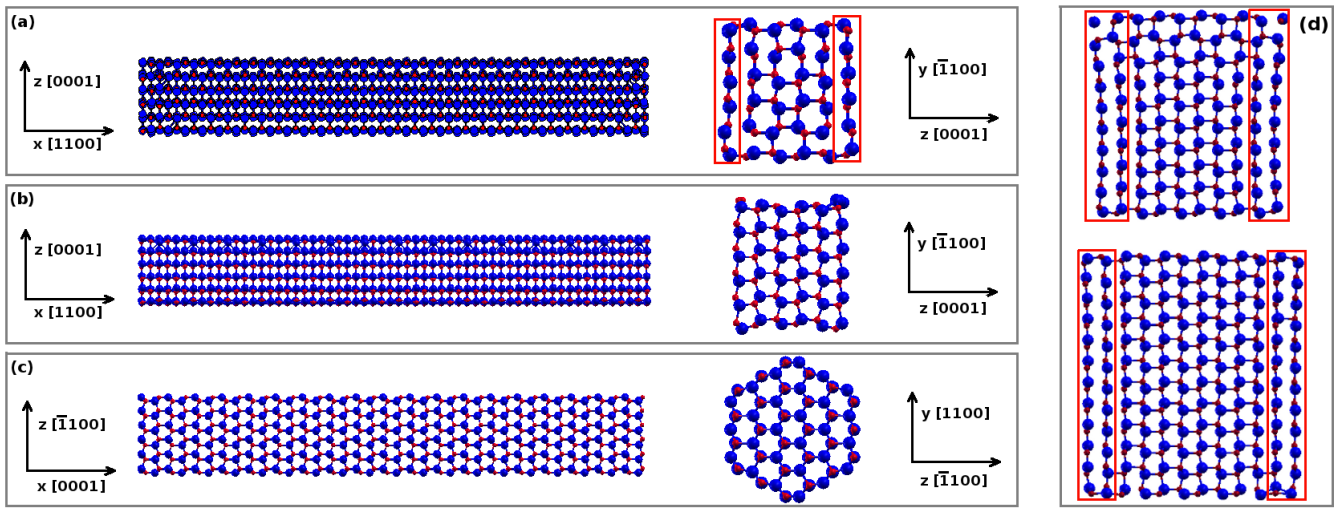}}
  \end{center}
  \caption{(Color online) The optimized configuration of three ZnO NWs studied in the present work. (a) Free NW: [1100]-oriented with free polar surfaces in the $z$ direction. Side view is on the left and the cross section is on the right. The free polar surfaces are reconstructed into shell-like structure (highlighted by red box). (b) Reduce NW: [1100]-oriented. The charges for ions on the polar surfaces are reduced by a factor of 0.25. (c) Hexagonal NW: [0001]-oriented without polar surface. The cross section is hexagonal. (d) Cross sections for two thicker free NWs, both showing the shell-like surface reconstruction at the free polar surfaces. The Zn and O ions are represented by big (blue online) and small balls (red online), respectively.}
  \label{fig_cfg}
\end{figure*}

Recently, the thermoelectric properties of ZnO NWs have received much attention due to the increasing global energy crisis.\cite{KulkarniAJ2006apl,IgamberdievK,ChantrenneP,ShiL2012,BuiCT2012small} The thermoelectric phenomenon can be utilized to harvest otherwise wasted thermal energy directly into electric power. To increase the thermoelectric efficiency, the thermal conductivity in NWs should be reduced, which can be realized through phonon confinement effects,\cite{BalandinA1998prb,KhitunA,KhitunA2000pe,ZouJ2001jap} isotope doping,\cite{YangN2008nl} surface roughness,\cite{MartinP,MartinP2010nl,LimJ} or non-planar (kinked) structures.\cite{JiangJW2013nl} Shi et al. suggested enhancing the thermoelectric efficiency through gallium doping, and the thermoelectric figure of merit was found to be increased by a factor of 2.5 at 4\% gallium doping.\cite{ShiL2012} In a recent experiment, Bui et al. measured the thermal conductivity of ZnO NWs with diameters in the 50-210~{nm} range.\cite{BuiCT2012small} A much lower thermal conductivity was observed as compared to bulk ZnO due to phonon scattering at the NW surface. The thermal conductivity was additionally found to decrease rapidly with decreasing NW thickness. These studies demonstrate the potential promise of thin ZnO NWs (which has low thermal conductivity) for thermoelectric applications.  However, as we have outlined above, surface effects dominate the physical properties of the ZnO NWs, especially when the surface of the NW is the (0001)-Zn or (000$\bar{1}$)-O polar surface. This polar surface effect on the thermal conductivity of the ZnO NW has not been investigated yet, and is thus the focus of the present work.

In this paper, we comparatively study the thermal conductivity in thin ZnO NWs under three different polar (0001) surface conditions using classical molecular dynamics (MD) simulations: with free polar surfaces, with reduced surface charges on the polar surfaces, and finally hexagonal NW without transverse polar surfaces.  We report significantly higher thermal conductivity in the ZnO NW with free polar surfaces. Specifically, the thermal conductivity in the ZnO NW with free polar surfaces is essentially double that of the hexagonal NW and the NW with reduced surface charges.  A normal mode analysis demonstrates that the underlying mechanism is a surface reconstruction to a shell-like structure that occurs on the free polar surfaces.  This shell-like surface reconstruction suppresses twisting motion in the NW, which prevents the bending phonon modes from being scattered by the other phonon modes.  Because the bending modes are preserved, they exhibit an extremely long lifetime, which leads to a significantly elevated thermal conductivity for ZnO NWs with free polar surfaces.  We further extract the phonon lifetime from the auto-correlation function, which explains the MD simulation results through a simple formula related to the phonon lifetime.

\section{structure and simulation details}

Our results on the thermal transport properties of ZnO NWs were obtained via classical MD simulations, where the interatomic interactions were described by the following Buckingham-type potential:
\begin{equation}
V_{\rm tot}(r_{ij}) = \sum_{i\not=j}^{N} A \exp(-\frac{r_{ij}}{\rho}) - \frac{C}{r^{6}_{ij}} + V_{\rm long}(r_{ij}),
\label{eq_buckingham}
\end{equation}
where $N$ is the total number of ions. The short-range parameters $A$, $C$, and $\rho$ for O were developed by Catlow,\cite{CatlowCRA} while the other short-range parameters are from Catlow and Lewis.\cite{LewisGV}  All ZnO NWs in our study have a hexagonal wurtzite crystal structure, where for bulk ZnO the lattice constant of the hexagonal lattice is $a=$3.2709~{\AA}; while the lattice constant in the perpendicular direction of the hexagonal plane is $c=$5.1386~{\AA}. The intra-cell geometrical parameter is $u=0.3882$.

Several efficient techniques have been developed to deal with the summation for the long-range electrostatic interaction between ions. The Ewald summation is the traditional way to calculate the electrostatic interaction within a bulk system with periodic boundary conditions.\cite{EwaldPP} However, for surface-dominated nanomaterials such as NWs in the present work, the periodicity assumption is clearly violated, and therefore it is crucial to employ a truncation-based summation method.  In this work, we have utilized the truncation-based summation approach initially proposed by Wolf et al. in 1999\cite{WolfD} and further developed by Fennell and Gezelter in 2006.\cite{FennellCJ}  A key development in the work by~\citet{FennellCJ} was to ensure that the electrostatic force and potential are consistent with each other, while remaining continuous at the interatomic cut-off distance.  In a recent study, Gdoutos et al. have quantified the errors in using the traditional periodic Ewald summation approach for ZnO NWs.\cite{GdoutosEE} In our calculation, we have chosen the damping parameter $\alpha=0.3$~{\AA$^{-1}$} and the cut-off $r_{c}=10.0$~{\AA}, which give convergent results for both the piezoelectric properties,\cite{DaiS2011jap} as well as the mechanical ZnO NWs.\cite{AgrawalR}

Fig.~\ref{fig_cfg} shows the relaxed, energy-minimized configuration of the three ZnO NWs we considered in this work, where the energy minimized configuration is obtained using the conjugate gradient method.  The NW in panel (a) is [1100]-oriented with dimension $98.1\times 17.0 \times 15.4$~{\AA}. The $y$ axis is along the $[\bar{1}100]$ direction, while the $z$ axis is along the [0001] polar direction. The side view is shown on the left while the cross section is on the right. The cross sectional view clearly shows a shell-like reconstruction at the two polar surfaces (highlighted by red boxes).  The Zn and O ions move into an essentially planar configuration which suppresses the surface polarization.\cite{DaiS2013jmps}  In this way, the outermost planes of atoms in the $z$ direction try to peel away from the bulk system, leading to shell-like layers on the (0001) and $(000\bar{1})$ surfaces. The shell-like reconstruction also happens in thicker ZnO NWs as shown in panel (d), where the cross sections are $28.3 \times 25.7$~{\AA} and $34.0 \times 30.8$~{\AA} for the two NWs, respectively. We will refer to the NW shown in panel (a) for the remainder of this work as the free ZnO NW.

The ZnO NW in panel (b) has the same orientation and cross sectional dimensions as the free ZnO NW in (a). The only difference is that the surface charges on the two polar surfaces are reduced by multiplying by a  factor of 0.75. More specifically, the charge of the Zn ions on the (0001)-Zn surface are reduced from $+2e$ to $+1.5e$, while the charge of the O ions on the $(000\bar{1})$-O surface are reduced from $-2e$ to $-1.5e$.  This charge reduction is justified by the fact that surface Zn and O atoms have only three nearest neighbors as compared to bulk Zn and O, which have four.  Hence, a 25\% reduction in the surface charges can be utilized to stabilize the polar (0001) surfaces.\cite{NoskerRW,NogueraC} A more rigorous physical and mathematical justification can be found in~\citet{NogueraC}.  In that work, from electrostatic and electronic structure calculations, it was found that a counterfield can be created to quench the macroscopic dipole moment if the Zn-terminated polar surface is less positive and the O-terminated polar surface is less negative by a factor of $R_{2}/(R_{1}+R_{2})\approx 0.75$, where $R_{1}=0.69$~{\AA} and $R_{2}=1.99$~{\AA} are the alternating distance for ion layers in the (0001) lattice direction.  It is clear from panel (b) that the crystal structure of the two polar surfaces is maintained after the energy minimization. In the following, we will refer to these NWs as reduce ZnO NWs.

Panel (c) shows the [0001]-oriented hexagonal ZnO NW. The length is 97.6~{\AA} and the two lateral dimensions are $19.6\times 17.0$~{\AA}. The [0001] polar direction is the axial direction and all side surfaces are neutral, so there are no transverse polar (0001) surfaces for these hexagonal ZnO NWs.  We will refer to such NWs as hexagonal ZnO NW in the following.

\begin{figure}[htpb]
  \begin{center}
    \scalebox{1.0}[1.0]{\includegraphics[width=8cm]{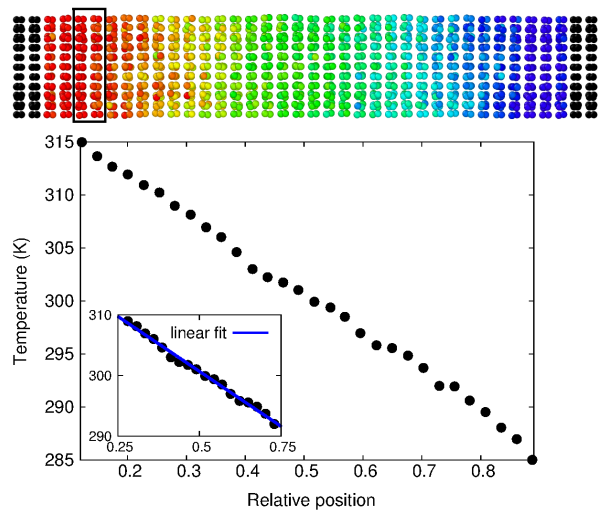}}
  \end{center}
  \caption{(Color online) The temperature profile from MD simulation for the hexagonal ZnO NW at 300~K. The $x$ axis is the relative axial position. Each point in the curve corresponds to the average temperature of atoms within the same atom layer shown in the top inset. Top inset: the three-dimensional kinetic energy distribution in the NW. Atoms are colored according to their average kinetic energy. There are 19 translational unit cells (highlighted by the black box), and each cell is divided into two atom layers. Bottom inset: the central 50\% region of the temperature profile is used to do linear fitting to extract the temperature gradient.}
  \label{fig_dTdx}
\end{figure}

It is important to note that the ZnO NW configurations we have chosen to study were done so because they are the most commonly synthesized and studied configurations experimentally.  First of all, the hexagon NW are commonly synthesized experimentally, particularly when ZnO NWs are grown vertically from a substrate, like sapphire, which possess six or three-fold symmetry.  Thus, the resulting epitaxially grown ZnO NWs prefer to have a hexagonal cross section.\cite{YangP2002} The free NWs are typically grown by sublimation of ZnO powder without introducing a catalyst.\cite{WangZL} An interesting phenomenon in the free NW is experimentally-observed planar defects parallel to the (0001) polar surfaces.\cite{DingY} The shell-like surface reconstruction in Fig.~\ref{fig_cfg}~(a) is also a planar defect and shares some similarities with this experimentally-observed planar defect. For the reduce NW, the charge reduction on the (0001) polar surface is one of the most promising mechanisms to stabilize the polar surface. This stabilization mechanism was proposed based on both electrostatic considerations\cite{NoskerRW} as well as electronic structure calculations\cite{ZwanzigR}.  Many experimental efforts have been devoted to investigate the actual process for this stabilization mechanism. For example, the charge reduction can be realized via the transfer of charges between (0001)-Zn and (000$\bar{1}$)-O polar surfaces (i.e., electronic relaxation).\cite{DulubO,LauritsenJV} 

In the present work, we focus on ZnO NWs of small (i.e. $<$ 5 nm$^{2}$) cross sectional size, for two main reasons.  First, recent experiments have found that the thermal conductivity in ZnO NWs decreases with decreasing cross-sectional area,\cite{BuiCT2012small} and so ultra-small NWs are more promising candidates for thermoelectric materials, which justifies the focus of the present work on ultra-small ZnO NWs.  Furthermore, calculation of the electrostatic term of the interatomic potential is fairly expensive computationally, which also limits the ability to study NWs with larger cross sectional dimensions.

Having described the surface treatments that are utilized, we now explain how the thermal transport properties were calculated.  As shown in the top inset of Fig.~\ref{fig_dTdx}, fixed boundary conditions are applied in the axial direction, i.e the two ends (black online) of the NW are fixed for the entire simulation, while free boundary conditions are applied in the transverse directions.  The Newtonian equations of motion are integrated using a velocity Verlet algoritm with a time step of 1.0 fs.  The total simulation time is typically about 2 ns, but can be extended to guarantee that a steady state condition is reached.  The two regions close to the left/right ends are maintained at high/low temperatures $T_{L/R}=(1\pm\alpha)T$ via a N\'ose-Hoover thermostat.\cite{Nose,Hoover} $T$ is the average temperature, and $\alpha=0.05$ is adopted in our calculation. Thermal energy is pumped into the system through the left temperature-controlled region with thermal current $J_{L}$, which will flow out of the right temperature-controlled region with current $J_{R}$. The energy conservation law requires $J_{L}=-J_{R}$ at steady state. Using this relation, the thermal current across the system can be calculated by $J=(J_{L}-J_{R})/2$, and the thermal conductivity is obtained from the Fourier law.

We note that we do not utilize a quantum correction in this work.  The quantum correction will become more important for temperatures below the Debye temperature.\cite{JiangJW2010isotopic}  Furthermore, experiments show that there is a maximum point in the temperature-dependence of the thermal conductivity around 130~K, which is insensitive to the diameter of the NW,\cite{BuiCT2012small} which indicates that the Debye temperature in the ZnO NW is around this temperature. As a result, the classical MD simulations in the present work give accurate values of thermal conductivity above 130~K, while the results will be overestimated for temperatures below 130~K.

\section{results and discussion}

\begin{figure}[htpb]
  \begin{center}
    \scalebox{1.0}[1.0]{\includegraphics[width=8cm]{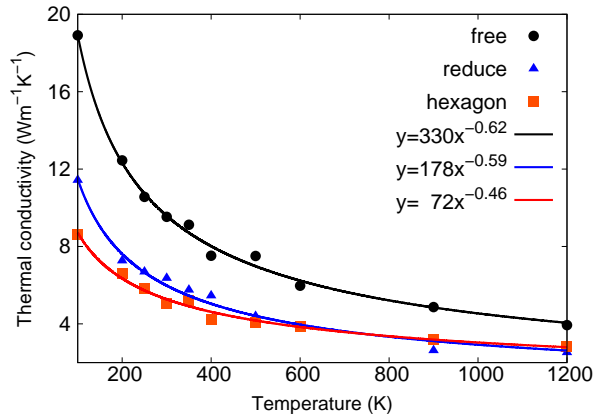}}
  \end{center}
  \caption{(Color online) The temperature dependence for the thermal conductivity in the free ZnO NW (black dots), the reduce NW (blue triangles), and the hexagonal NW (red squares).}
  \label{fig_kappa}
\end{figure}

Fig.~\ref{fig_dTdx} shows the temperature profile from MD simulation for the hexagonal ZnO NW at 300~K. The $x$ axis in the figure is the relative axial position. Each point in the curve is obtained by averaging the temperature of atoms within the same atom layer shown in the top inset. The top inset displays the three-dimensional kinetic energy distribution within the NW. Atoms are colored according to their average kinetic energy. There are 19 translational unit cells (highlighted by the black box) in the hexagonal NW, and each unit cell contains two atom layers. We have used the central 50\% region of the temperature profile to do linear fitting to extract the temperature gradient across the NW (see bottom inset in the figure).

We calculate the thermal conductivity for the three ZnO NWs (free, reduce, hexagonal) defined previously. The temperature dependence for the thermal conductivity is shown in Fig.~\ref{fig_kappa} for the free NW (black points), the reduce NW (gray pentagons), and the hexagonal NW (red squares).  For all three NWs, the thermal conductivity decreases with increasing temperature for the full temperature range. This result can be understood as follows. In classical MD simulations as employed here, all phonon modes are sufficiently excited even at low temperatures, i.e each phonon mode contributes $k_{B}/2$ to the heat capacity. Hence, the temperature dependence in the thermal conductivity comes from temperature dependence of the phonon lifetime. The phonon lifetime is limited by the phonon-phonon scattering in the simulation, which becomes stronger at higher temperature. As a result, the phonon lifetime is shorter at higher temperature, leading to a decrease in the thermal conductivity with increasing temperature for the whole temperature range. This is different from experiments, where the thermal conductivity increases with increasing temperature for temperatures below the Debye temperature (130~K) due to quantum effects.\cite{BuiCT2012small}

\begin{figure}[htpb]
  \begin{center}
    \scalebox{1.1}[1.1]{\includegraphics[width=8cm]{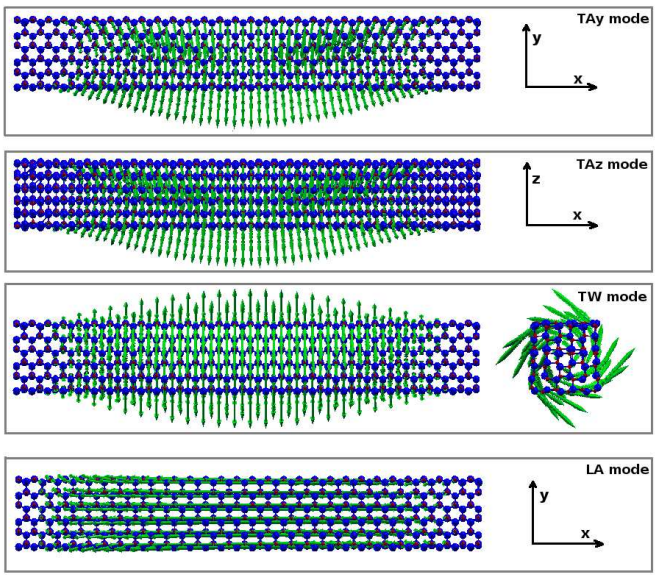}}
  \end{center}
  \caption{(Color online) The polarization vector of the four acoustic phonon modes in the free ZnO NW. From top to bottom: TA$_{y}$, TA$_{z}$, TW, and LA modes. Arrows in the figure represent the polarization vector of the phonon mode. The arrows for the LA mode are in the $x$ direction.}
  \label{fig_u}
\end{figure}

Fig.~\ref{fig_kappa} shows that the thermal conductivity decreases as an exponential function $T^{-\alpha}$, with exponent $\alpha=$ 0.62, 0.59, and 0.46 for the free, reduce, and hexagonal ZnO NWs, respectively. These exponents from our simulations are smaller than 1.0, which likely is due to the strong surface scattering for the phonon modes. The surface or boundary scattering is a common mechanism in low-dimensional NW structures, which is closely related to weaker phonon-phonon scattering.\cite{BalandinAA2008,NikaDL2009prb,NikaDL2009apl,Balandin2011nm} The experimental value for the exponent is above 1.0, which is attributed to other phonon scattering mechanisms besides the phonon-phonon scattering,\cite{BuiCT2012small} for example intrinsic vacancy point defects or isotropic disorder. In our simulations, all ZnO NWs are pristine without defects, so the temperature exponents from our calculations are smaller than the experimental values.  For similar reasons, the thermal conductivity in our study is obviously higher than experimentally measured values of thermal conductivity for ZnO NWs of similar cross sectional area, which can be extracted from the diameter dependence for the thermal conductivity measured in the experiment.\cite{BuiCT2012small}

The reduce and hexagonal ZnO NWs have similar thermal values for thermal conductivity, especially in the high temperature region where their thermal conductivities are almost the same. In low temperature region, the thermal conductivity in the reduce NW is slightly higher than that of the hexagonal NW. One distinct phenomenon in Fig.~\ref{fig_kappa} is that the thermal conductivity in the free ZnO NW is basically double that of the reduce or hexagonal NWs for the full temperature range.
\begin{figure}[htpb]
  \begin{center}
    \scalebox{1.1}[1.1]{\includegraphics[width=8cm]{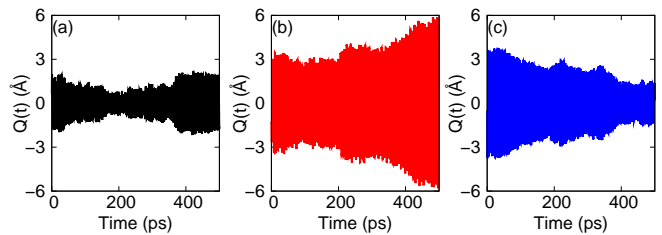}}
  \end{center}
  \caption{(Color online) The normal mode coordinate at room temperature for the TW mode in the free ZnO NW (left), the reduce NW (middle), and the hexagonal NW (right).}
  \label{fig_Qt_300K_TW}
\end{figure}
It should be noted that the thermal conductivity in bulk ZnO is slightly anisotropic due to its hexagonal wurtzite lattice structure.\cite{WolfMW1973pssa} The anisotropic effect leads to about 15\% difference in the thermal conductivity along [1100] or [0001] lattice directions. The difference between the thermal conductivity in the free and hexagonal ZnO NWs is much larger than the anisotropic effect. Hence, the present difference should be attributed to other effects. Different from the anisotropic effect, surface effects plays a dominant role in the thermal transport in NWs with large aspect of ratio. To reveal the importance of the surface effects, we calculate the thermal conductivity of a [1100]-oriented NW with the same structure as the reduce ZnO NW except that periodic boundary conditions are applied in the two transverse directions. The relaxed structure for the NW with periodic boundaries is similar as the reduce NW, and we find that the thermal conductivity is 15.09~{Wm$^{-1}$K$^{-1}$} at 300~K for this periodic boundary condition case. This value is about two times larger than the value of 6.36~{Wm$^{-1}$K$^{-1}$} in the reduce NW. It is also larger than the thermal conductivity at 300~K in the free NW (9.53~{Wm$^{-1}$K$^{-1}$}) and hexagon NW (5.05~{Wm$^{-1}$K$^{-1}$}). This result illustrates the importance of the finite-size effect and surface effect on the thermal conductivity of the ZnO NW. Hence, it is natural to conclude that the high thermal conductivity in the free NW is due to its special surface configuration.

A novel feature in the free NW is the shell-like surfaces resulting from the surface reconstruction in the $z$ direction, whereas there is no shell-like reconstruction in the other two NWs.  Because of this, our key insight is that the shell-like surface structure may act as a safeguard for the bending movement in the free ZnO NW. More specifically, the shell-like surfaces prevent the ZnO NW from exhibiting torsional motion during the thermal transport process. In other words, the bending modes in the ZnO NW can efficiently deliver heat energy across the system without being scattered by the twisting modes. As a result, the bending modes have a much longer lifetime in the free ZnO NW, which directly results in higher thermal conductivity. This preserving mechanism for the bending mode is crucial for the much higher thermal conductivity of the free ZnO NW because it is well-known that thermal energy is mainly transported by the bending modes in both two-dimensional and one-dimensional nanosystems.\cite{NikaDL2009prb,NikaDL2009apl,LindsayL} Hence, the preservation of the bending mode directly leads to much higher thermal conductivity in the free ZnO NW.

To verify the preserving mechanism, we calculate the normal mode coordinate via the following three steps.
\begin{enumerate}
\item Prior to conducting the MD simulations, we derive the dynamical matrix from the interatomic potential $V$ through the formula
\begin{equation}
K_{ij}=\partial^{2}V/\partial x_{i}\partial x_{j},
\label{eq_Kij}
\end{equation}
where $x_{i}$ is the position of the $i$-th degree of freedom. This formula is realized numerically by calculating the energy change after a small displacement of the $i$-th and $j$-th degrees of freedom.
\item The dynamical matrix is diagonalized, resulting in the eigenvalue $\omega_{k}$ (i.e., phonon frequency) and the eigenvector {\boldmath $\xi$}$^{k}=(\xi^{k}_{1}, \xi^{k}_{2}, \xi^{k}_{3}, ..., \xi^{k}_{3N})$. This eigenvector is used as a projector.  Specifically, the operation of {\boldmath $\xi$}$^{k}$ on a displacement vector projects the vector onto the motion style corresponding to a particular normal mode $k$. Both ends of the NW are fixed in this eigenvalue calculation, which is consistent with the fixed boundary condition applied in the above thermal transport simulation. 
\item During the MD simulation, we use the projector constructed by the eigenvector to extract the motion style of a particular normal mode, such as the twisting mode, the bending mode, etc. This projection process is actually the calculation of the time history for the normal mode, i.e the normal mode coordinate, as follows,
\begin{equation}
Q^{k}(t)=\sum_{j=1}^{3N}\xi^{k}_{j}(r_{j}(t)-r_{j}^{0}),
\label{eq_Qt}
\end{equation}
where $k$ is the mode index, $3N$ is the total number of degrees of freedom in the NW, $r_{j}(t)$ is the trajectory of the $j$-th degree of freedom from the MD simulation and $r_{j}^{0}$ is the corresponding equilibrium position.
\end{enumerate}
Fig.~\ref{fig_u} displays the polarization vector of the four lowest-frequency acoustic phonon modes in the free NW. From top to bottom, the four phonon modes are the $y$ transverse acoustic (TA$_{y}$) mode, the $z$ transverse acoustic (TA$_{z}$) mode, the twisting (TW) mode, and the longitudinal acoustic (LA) mode. The TA$_{y}$ and TA$_{z}$ are the two bending modes.

\begin{figure}[htpb]
  \begin{center}
    \scalebox{1.1}[1.1]{\includegraphics[width=8cm]{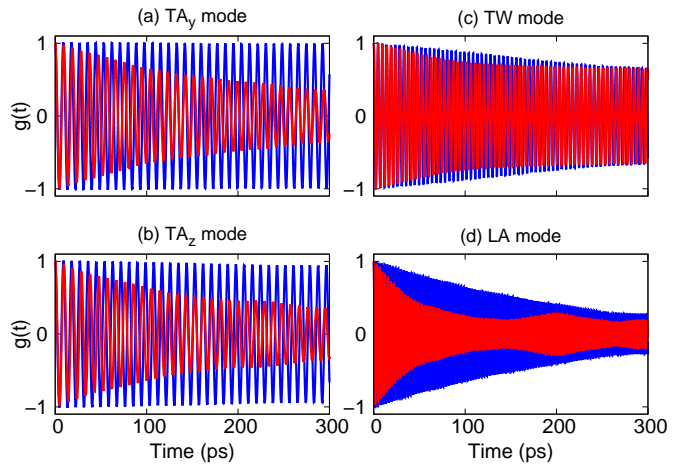}}
  \end{center}
  \caption{(Color online) The auto-correlation function $g(t)$ for the four acoustic phonon modes in the free ZnO NW at 300~K (blue online) and 1200~K (red online), where the curves for 1200~K are plotted on top of those at 300~K. (a) The TA$_{y}$ mode. (b) The TA$_{z}$ mode. (c) The TW mode. (d) The LA mode.}
  \label{fig_gt_free}
\end{figure}
\begin{figure*}[htpb]
  \begin{center}
    \scalebox{1.0}[1.0]{\includegraphics[width=\textwidth]{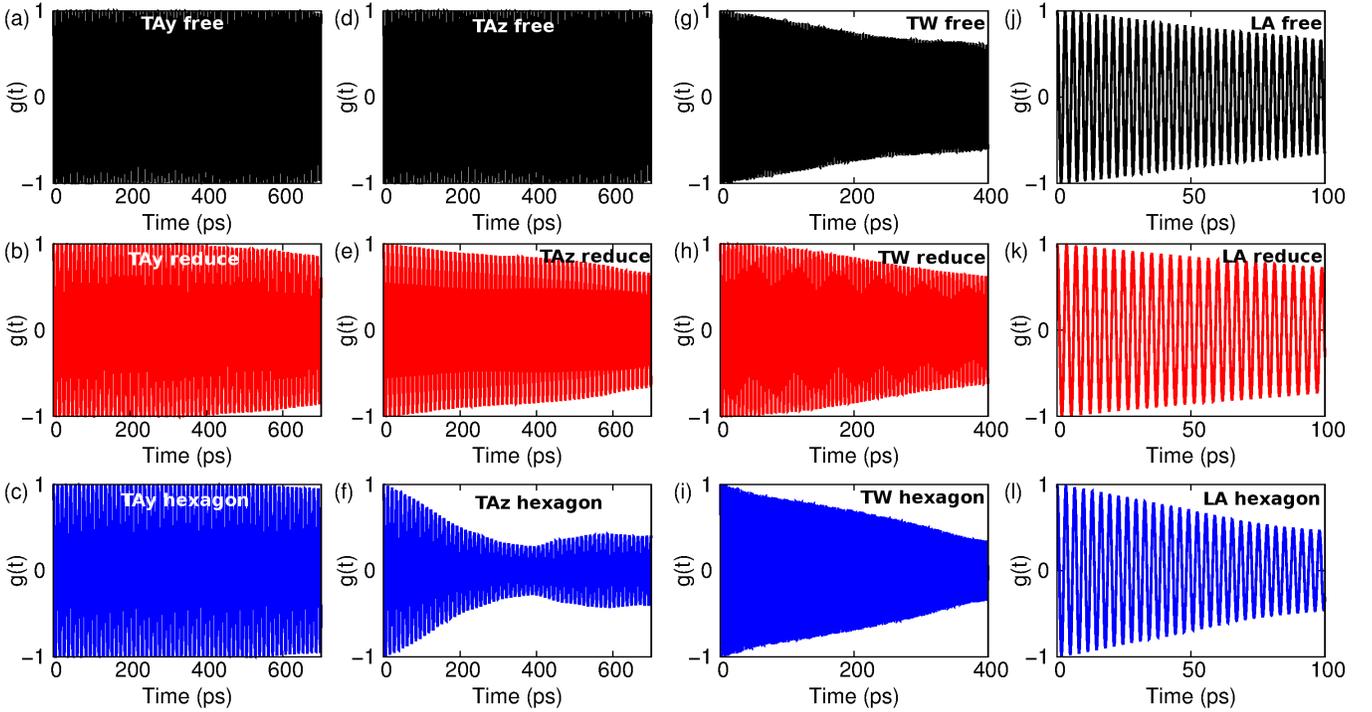}}
  \end{center}
  \caption{(Color online) The auto-correlation function $g(t)$ at room temperature for the four acoustic phonon modes in the three ZnO NWs. (a)-(c) are for the TA$_{y}$ modes in the three NWs. (d)-(f) are for the TA$_{z}$ modes. Note obvious decay in the correlation function in (e) and (f) for the reduce and hexagonal NWs. (g)-(i) are for the TW modes. (j)-(l) are for the LA modes.}
  \label{fig_gt_300K}
\end{figure*}

The four polarization vectors of the four phonon modes in Fig.~\ref{fig_u} describe four particular vibrational modes of the free ZnO NW. From an energetic point of view, these four vibrational modes are the most favorable modes of motion in the NW as they have the lowest frequency, which is equivalent to saying they require the least amount of energy to excite.  The normal mode coordinate for a particular vibrational mode is useful as it elucidates the relative importance of a given vibrational mode to the overall motion of the NW.  If the amplitude of the normal mode coordinate for a particular phonon mode is large, it means that the corresponding vibrational mode is energetically favorable for the NW.  The time history of the normal mode coordinate at room temperature for the TW modes are shown in Fig.~\ref{fig_Qt_300K_TW} for (a) the free ZnO NW, (b) the reduce ZnO NW, and (c) the hexagonal ZnO NW. Indeed, we find that the $Q(t)$ for the TW mode in the free ZnO NW is much smaller than in the other two ZnO NWs. This result means that the twisting movement is the weakest in the free ZnO NW, which provides initial evidence for our hypothesis that the shell-like surfaces in the free NW are indeed able to prevent the NW from exhibiting large torsional motion.

It should be pointed out that our MD simulations were performed on a finite structure (i.e. not a bulk, periodic system) and that the two ends of the NW are fixed during the whole simulation process.  Periodicity is clearly absent in these one-dimensional NWs, and as a result the wave vector $\vec{k}$ lacks a proper definition, as the wave vector is traditionally defined with respect to the translational symmetric operation on a unit cell. If the entire NW in our study is regarded as a big `unit cell', then the `wave vector' is zero for all normal modes in this situation. A proper definition for the mode index $k$ in our manuscript is the number index of the normal mode according to its frequency order. That is, all normal modes are ordered with respect to their frequency, and the mode index $k=1,2,3,...,3N$. In our study, we have analyzed four acoustic modes with the lowest frequency: the first $TA_{y}$ mode (i.e., this mode has the lowest-frequency among all $TA_{y}$ modes), the first $TA_{z}$ mode, the first TW mode, and the first LA mode. We have chosen these four representative modes, because the acoustic phonon modes, and especially the low-frequency modes, make largest contribution to the thermal conductivity in NWs.

We further verify that the weak twisting in the free ZnO NW can preserve the bending modes from being scattered, and thus yields a long lifetime for the bending modes. To this end, we calculate the auto-correlation function of the above normal mode coordinate. The normalized auto-correlation function $g(t)$ is calculated by,
\begin{eqnarray}
g^{k}(t)=\frac{\langle Q^{k}(t)Q^{k}(0)^{*}\rangle}{\langle Q^{k}(0)Q^{k}(0)^{*}\rangle}.
\label{eq_gt}
\end{eqnarray}
For harmonic systems without phonon-phonon scattering, this equation simply gives $g(t)=\cos \omega t$; i.e the oscillator vibrates without decay, indicating an infinite lifetime. In the MD simulation, the phonon-phonon scattering reduces the phonon lifetime to be finite. Under the single mode relaxation time approximation,\cite{LaddAJC,McgaugheyAJH} we have $g^{k}(t)=\cos \omega_{k} t e^{-t/\tau_{k}}$, where $\omega_{k}$ is the angular frequency of the phonon mode and $\tau_{k}$ is the lifetime. With this approximation the depopulation of the phonon states is described rigorously, but it fails on the corresponding repopulation. The latter is assumed to have no memory of the initial phonon distribution.

Fig.~\ref{fig_gt_free} shows the auto-correlation function for the four acoustic phonon modes in the free ZnO NW at 300 and 1200~K. A distinct phenomenon is that there is almost no decay in the auto-correlation functions of the two bending modes at 300~K, which implies extremely long lifetime for these two bending modes, and is the reason why the bending modes make the most important contribution to the thermal transport in the low-dimensional nanomaterials. For instance, theoretical calculations have revealed a superior lifetime of the bending modes in two-dimensional graphene.\cite{NikaDL2009prb,NikaDL2009apl,LindsayL} Our auto-correlation function calculations are consistent with these results. The TW and LA modes have much shorter lifetime than the bending modes at 300~K. From the comparison between these curves for 300K and 1200~K, we find that the lifetime for the two bending modes and the LA mode decrease considerably with increasing temperature. However, the lifetime of the TW mode at 1200~K is almost the same as that at 300~K. Hence, the twisting mode makes a major contribution to the thermal transport in the free ZnO NW at high temperature. 

\begin{table*}
\caption{The lifetime $\tau$ (in ps) and the angular frequency $\omega$ (in 1/ps) of the four acoustic phonon modes under different conditions. The angular frequency is in the parenthesis. $\tau$ is obtained by fitting the correlation function to $g(t)=\cos (\omega t) e^{-t/\tau}$, where $\omega$ is the angular frequency of the phonon mode. Note that the lifetime for the TA$_{z}$ modes in the reduce and hexagon NWs are obviously shorter than that in the free NW (in the 3rd column).}
\label{tab_lifetime}
\begin{tabular*}{\textwidth}{@{\extracolsep{\fill}}|c|c|c|c|c|}
\hline
{\backslashbox{condition}{mode}} & TA$_{y}$ & TA$_{z}$ & TW & LA \\
\hline
free 1200 K & 240.0 (0.66) & 221.6 (0.66) & 298.3 (1.30) & 62.2 (2.02)\\
\hline
free 300 K & 8215.31 (0.71) & \textcolor{blue}{7545.0} (0.70) & 673.4 (1.33) & 230.0 (2.11)\\
\hline
reduce 300 K & 4730.5 (0.56) & \textcolor{blue}{2026.4} (0.54) & 918.7 (1.03) & 287.9 (1.78)\\
\hline
hexagon 300 K & 6712.6 (0.55) & \textcolor{blue}{307.6} (0.56) & 497.1 (1.39) & 126.9 (1.92)\\
\hline
\end{tabular*}
\end{table*}
We note a recurrence phenomenon in the auto-correlation function of the LA mode at 1200~K, where the amplitude of the auto-correlation function shows an increase around 200~{ps}. The role of this effect should be reasonably small, as long as it happens after the complete decay of the auto-correlation function, and we have accordingly used simulation data before the onset of the recurrence phenomenon. The understanding for this phenomenon is still lacking. However, we point out that it should not be the result of the phonon reflection at the fixed ends,\cite{ZwanzigR} which should happen at a recurrence time around 2.0~ps.  No evidence of the recurrence phenomenon around 2.0~ps is observed for any of the auto-correlation functions.  This phonon reflection effect has been avoided due to the incorporation of a modified fixed boundary condition that has been applied in our simulation. That is, we have slightly shifted the heat bath away from the two fixed ends. As shown in the inset of Fig.~\ref{fig_dTdx}, two atomic layers are used to separate the fixed region and the heat bath-controlled region. We have shown that this modified fixed boundary condition succeeds in avoiding boundary temperature jumps by reducing the energy localization of the edge mode.\cite{JiangJW2009edge} The separator space also is crucial to avoiding the phonon reflection effect because the phonon modes need to pass through the heat bath twice (i.e. once moving towards the fixed end and once moving back towards the heat bath-controlled region) before coming back to the center region of the NW.  The effect of this is that the phonon modes will effectively be burnt off during the two trips through the heat bath region. 

A more rigorous explanation is based on the localized edge mode, which is the description of the phonon reflection effect within the lattice dynamics concept.\cite{JiangJW2009edge} We have previously found that there are some edge phonon modes at the two ends of the nano-material.\cite{JiangJW2009edge} These edge modes are localized in the edge region, i.e only edge atoms have large vibrational amplitudes in these modes. If the heat bath is applied to the edge region, the localized edge modes will be sufficiently excited. As a result, the thermal energy will be localized in the edge region, due to the special vibration morphology of the edge modes. Corresponding to this localization effect, a large temperature jump will occur at the boundary region. After shifting the heat bath slightly away from the edge region, the edge modes are very difficult to be excited, because atoms in the heat bath have almost no vibrational amplitude in the eigenvector of the edge mode. As a result, the edge effect is suppressed and the boundary temperature jump is removed. Fig.~\ref{fig_dTdx} shows a clean temperature profile without any obvious temperature jumps at the two boundaries, which reflects the reduction of the phonon reflection effect.

Fig.~\ref{fig_gt_300K} shows the correlation function $g(t)$ at room temperature for the four acoustic modes in the three ZnO NWs. Panels (a)-(c) show that the TA$_{y}$ bending mode in the free NW has the longest lifetime among the three studied NWs. The lifetime of this bending mode are slightly decreased in the reduce and hexagonal NWs. Panels (d)-(f) reveal obvious decay in the auto-correlation function for the reduce and hexagonal NWs, which illustrates that the lifetime of the TA$_{z}$ bending mode in the free NW is much longer than the lifetime of the TA$_{z}$ bending modes in the reduce and hexagonal NWs due to the shell-like surface reconstruction of the polar (0001) surfaces of the free ZnO NW. The recurrence phenomenon is also observed in panel (f). Panels (g)-(i) are for the TW modes, and panels (j)-(l) are for the LA modes. The lifetime of the TW and LA modes are close to each other in all of the three NWs.

Tab.~\ref{tab_lifetime} summarizes all phonon lifetimes, which are obtained by fitting the auto-correlation to $g^{k}(t)=\cos \omega_{k} t e^{-t/\tau_{k}}$. Special attention should be given to the second and the third columns for the two bending modes in the different NWs. Both bending modes have very long lifetimes in the free ZnO NW. For the reduce ZnO NW, the lifetime for both bending modes is considerably reduced. In the hexagonal NW, the lifetime of the TA$_{y}$ bending mode is only slightly decreased, but the lifetime of the TA$_{z}$ bending mode is substantially reduced. Consequently, as we have observed in above MD simulations, the thermal conductivity in the free ZnO NW is much higher than that in the reduce and hexagonal NWs.

Using Tab.~\ref{tab_lifetime}, we can propose a concise interpretation for our observations based on the MD simulations that the thermal conductivity in the free ZnO NW is essentially double that of the other two NWs, while the reduce and hexagonal ZnO NWs have almost the same thermal conductivity. The kinetic theorem says that the thermal conductivity ($\kappa$) is proportional to the lifetime of the phonon mode,\cite{HoneJ} i.e $\kappa\propto Cv^{2}\tau$, where $C$ is the heat capacity and $v$ is the phonon group velocity. Tab.~\ref{tab_lifetime} shows that the lifetimes of the two TA modes are about one order of magnitude higher than the TW and LA modes. Furthermore, the frequencies of the two TA modes are only about one half of the TW and LA modes, so the phonon density of state is much higher for the TA modes than the TW and LA modes. Consequently, the two TA modes make the most important contribution to the thermal conductivity. This is similar as what have been observed in both graphene and carbon nanotubes, where the TA modes dominate the thermal conductivity.\cite{NikaDL2009prb,JiangJW2011bntube} From this qualitative analysis, the contribution from the TW and LA modes should be at least one order of magnitude smaller than the TA modes, and forms the justification for neglecting the contribution from the TW and LA modes in the present analysis.  Overall, the error due to the neglect of the LA and TW modes should be less than 10\% in this analysis.  

We find that the thermal conductivity is proportional to the summation of the lifetime of the two bending modes, i.e $\kappa\propto ( \tau_{TA_{y}} + \tau_{TA_{z}} )$, with the assumption of only small variations in the heat capacity and the phonon group velocity for the bending modes in the three NWs. From the second and the third columns in the table, the summation $(\tau_{TA_{y}} + \tau_{TA_{z}}) = $ 15760.3, 6756.9, and 7020.2 ps for the free, reduce, and hexagonal ZnO NWs, respectively. It is quite interesting that the value of $(\tau_{TA_{y}} + \tau_{TA_{z}})$ in the free ZnO NW is roughly twice of that in the other NWs, while this value is almost the same in the reduce and hexagonal NWs.  The ordering of this value is the same as the ordering of the thermal conductivity values for these NWs.  Thus, this simple mathematics further supports our finding that shell-like reconstruction is the fundamental mechanism for the distinctly higher thermal conductivity in the free ZnO NW.

Finally, we note an interesting property trade-off relating to the polar surface reconstruction.  Through our MD simulations, we find that the reconstruction of the polar (0001) surface is beneficial for thermal transport properties. However, it has been found that this surface reconstruction degrades the piezoelectric properties.\cite{DaiS2013jmps} As a result, the decision of whether to use free polar surfaces should depend on the specific application of interest.

\section{conclusion}
In conclusion, we have performed MD simulations to comparatively study the thermal transport in three different ZnO NWs:  those with free polar (0001) surfaces, those with reduced surface charges to stabilize the polar (0001) surfaces, and hexagonal cross section NWs that do not have any transverse polar (0001) surfaces.  We find that the thermal conductivity in the free ZnO NW is much higher than the thermal conductivity in the other two NWs. By comparing the normal mode coordinates of the four acoustic modes, we find that the free polar surfaces undergo a shell-like reconstruction, which prevents the bending modes in the free NW from being scattered by the TW modes. We further analyze the lifetime of the four acoustic modes and find that the summation of the lifetime for the two bending modes in the free ZnO NW is about double that of the other two NWs, which gives a direct explanation for the highest thermal conductivity in the free ZnO NW.

\textbf{Acknowledgements} The work is supported by the German Research Foundation (DFG).  HSP acknowledges support from the Mechanical Engineering Department of Boston University.

%\bibliographystyle{apsrev4-1}
%\bibliography{biball}
\end{document}